\documentclass[twoside]{article}
\usepackage{spcof_icme,amssymb,amsmath,epsfig}
\usepackage{multirow}
\usepackage{hyperref}
\ninept  

\title{A Study on Incorporating Whisper for Robust Speech Assessment}
\name{Ryandhimas E. Zezario$^1$$^2$, Yu-Wen Chen$^3$, Szu-Wei Fu$^4$, Yu Tsao$^2$, Hsin-Min Wang$^2$, Chiou-Shann Fuh$^1$}
\address{
  $^1$National Taiwan University
  $^2$Academia Sinica 
  $^3$Columbia University
  $^4$NVIDIA   
  }

\begin{document}
\maketitle
\begin{abstract}
This research introduces an enhanced version of the multi-objective speech assessment model--MOSA-Net+, by leveraging the acoustic features from Whisper, a large-scaled weakly supervised model. We first investigate the effectiveness of Whisper in deploying a more robust speech assessment model. After that, we explore combining representations from Whisper and SSL models.  The experimental results reveal that Whisper's embedding features can contribute to more accurate prediction performance. Moreover, combining the embedding features from Whisper and SSL models only leads to marginal improvement. As compared to intrusive methods, MOSA-Net, and other SSL-based speech assessment models, MOSA-Net+ yields notable improvements in estimating subjective quality and intelligibility scores across all evaluation metrics in Taiwan Mandarin Hearing In Noise test - Quality \& Intelligibility (TMHINT-QI) dataset. To further validate its robustness, MOSA-Net+ was tested in the noisy-and-enhanced track of the VoiceMOS Challenge 2023, where it obtained the top-ranked performance among nine systems.

\end{abstract}
\section{Introduction}
Effectively evaluating the quality and intelligibility of spoken audio plays a critical role in many speech-related applications. One of the straightforward approaches to conducting the evaluation is by using a listening test based on human participants. In order to obtain human label scores, which are also known as subjective scores, a group of people are asked to listen to speech samples and provide feedback on their quality or intelligibility levels. The mean opinion score (MOS) is a common numerical indicator used in listening tests to assess speech quality, with a scale of one to five, where higher scores indicate better quality. On the other hand, the intelligibility score is based on the number of words, phonemes, or sentences correctly recognized in the played speech samples. To obtain a fair evaluation result, a large number of listeners are necessary, which can be costly and decrease practicality. To address this problem, various objective metrics for assessing quality and intelligibility based on signal processing techniques have been proposed, for instance, perceptual evaluation of speech quality (PESQ) \cite{ref_19}, perceptual objective listening quality analysis (POLQA) \cite{polqa_2013}, speech transmission index (STI) \cite{ref_38}, normalized-covariance measure (NCM) \cite{ncm}, short-time objective intelligibility (STOI) \cite{ref_39}, extended STOI (eSTOI) \cite{estoi}, spectrogram orthogonal polynomial measure (SOPM) \cite{somr}, neurogram orthogonal polynomial measure (NOPM)\cite{nopm}, and neurogram similarity index measure (NSIM) \cite{nsim}. Although signal processing-based objective metrics for speech assessment have performed well, obtaining a clean reference is typically necessary while performing the evaluation. Therefore, based on such concerns, researchers started to employ deep learning models for deploying non-intrusive speech assessment metrics that do not require clean reference.  

Deep learning-based speech assessment metrics can be categorized into two classes, depending on whether the ground-truth scores are obtained through subjective listening test or from a particular objective metric. More specifically, the first category of approaches is to predict human subjective ratings \cite{mbnet_mos, wav2vec_mos, ldnet2022, ssl-mos, mosa-net}, and the second category is to predict objective evaluation scores \cite{mosa-net, ref_49, ref_52, ref_56}. Compared to objective evaluation scores, predicting subjective assessment scores is more challenging, as each listener brings their own bias. Previous studies \cite{mbnet_mos,ldnet2022, Cooper2024e24} have suggested that accounting for individual listener information can improve prediction performance. Furthermore, recent advances in discriminative self-supervised learning (SSL) models have shown promising results when combined with speech assessment models as an additional module \cite{ssl-mos, yang22o_interspeech, saeki22c_interspeech} or used as a feature extractor \cite{mosa-net,zezario2022mti,zezario2022mbi}, leading to significant improvements in prediction accuracy.

Recently,  Whisper \cite{Whisper}, a large pre-trained model based on weak supervision, has been proposed and shown to have good potential for generating more robust acoustic features. This is due to the availability of audio transcripts in different languages and tasks. Unlike the SSL model that predicts the masked audio, the weak supervision model uses the actual transcript during model training. Because of this, the audio features generated by Whisper are expected to contain more phonetic information. Hence, it is worth investigating whether Whisper can provide more informative features for the speech assessment task.

In this study, we aim to explore the potential advantages of speech representations from Whisper, and propose an improved version of the multi-objective speech assessment model, namely MOSA-Net+. MOSA-Net+ incorporates three distinct features: traditional spectral features, waveforms processed using adaptable filters from a convolutional network \cite{sincnet}, and latent representations obtained from Whisper. Within our framework, the pre-trained Whisper module is accompanied by an additional adapter layer, facilitating task-specific adaptation and dimension reduction. MOSA-Net+ employs a multitasking learning approach to predict subjective quality and intelligibility scores. Its architecture comprises a convolutional neural network (CNN), followed by a bidirectional long short-term memory (BLSTM) and fully connected layers. Each task-specific layer includes an attention mechanism, a fully connected layer, and a global average pooling to obtain an estimated utterance score. 

The contribution of this study is twofold; first, we investigate the effectiveness of using the speech representations from Whisper in deploying a speech assessment model. Second, we explore the potential advantages of combining the embedding features from Whisper and SSL models while deploying MOSA-Net+. Experimental results in  Taiwan Mandarin Hearing In Noise test - Quality \& Intelligibility (TMHINT-QI) \cite{TMINT-QI} dataset first confirmed that Whisper embedding features can improve prediction performance for deploying the MOSA-Net+ model. Second, combining Whisper and SSL embedding can improve performance, but the improvement is rather marginal. Meanwhile, MOSA-Net+ notably outperforms several intrusive methods, MOSA-Net, and the other SSL-based assessment models in estimating subjective quality and intelligibility scores across all evaluation metrics, confirming Whisper's potential to provide more robust acoustic features \cite{gong2023whisper}. In order to further validate its performance, MOSA-Net+ was evaluated in the noisy-and-enhanced track of the VoiceMOS Challenge 2023 \cite{cooper2023voicemos}, emerging as the top-performing model among nine systems.

The rest of this paper is organized as follows. Section II presents the proposed MOSA-Net+. Section III explains Whisper and SSL's embedding analysis. Section IV describes the experimental setup and result. Finally, the conclusions and future work are presented in Section V. 

\section{MOSA-Net+}
\subsection{Architecture}
The MOSA-Net+ model's overall architecture is presented in Fig. 1. This model employs cross-domain acoustic features to predict multiple assessment scores. To process a speech waveform $\bold{Y}$, the model takes two input branches. In the first branch, the waveform is processed using Short-Time Fourier Transform (STFT) and learnable filter banks (LFB) of the convolutional network \cite{sincnet} separately. The resulting power spectral (PS) and LFB features with a time dimension $T$ and feature dimension $F$ are then concatenated and fed into a convolutional layer as follows:
\begin{equation}
 \begin{array}{c}
 \bold{PS} = STFT(\bold{Y}) 
 \\
 \bold{LFB} = Sinc Conv(\bold{Y}) \\
 \bold{Concat} = [\bold{PS}  |  \bold{LFB}] \\
 \bold{Conv} = CNN (\bold{Concat}) \\
 \end{array} 
\label{eq:eq1}
\end{equation}
In the second branch, the waveform undergoes processing via Whisper to generate embedding features, referred to as WS. This Whisper module is set to be frozen during model training. Next, these extracted WS features then undergo further processing through an adapter layer, facilitating task-specific adaptation and dimensional reduction before being concatenated in the following sequence:

\begin{equation}
 \begin{array}{c}
 \bold{WS} = Whisper(\bold{Y}) 
 \\
 \bold{WS}_{adapter} = Adapter(\bold{WS}) \\
 \bold{Concat}_{WS} = [\bold{Conv}  | \bold{WS}_{adapter}] \\
 \end{array} 
\label{eq:eq1}
\end{equation}
It is noteworthy that different types of features are concatenated by temporal dimension. Specifically, the number of frames in an utterance is the sum of frame numbers from the PS, LFB, and WS features. The combined features ($Concat_{WS}$) undergo processing through a bidirectional layer and a fully connected layer. Task-specific layers are employed to predict speech assessment metrics, utilizing attention mechanisms to focus on the more important regions. Subsequently, a fully connected layer is used for each metric to derive frame-level scores. The frame-level scores are aggregated using a global average operation to obtain the predicted Quality and Intelligibility scores. Finally, MOSA-Net+ integrates both frame-level and utterance-level scores into the objective function \cite{mosa-net}.

\begin{equation}
\label{eq:loss}
   \small
    \begin{array}{c}
   L_{All} = \gamma_{1}L_{Quality} + \gamma_{2}L_{Intelligibility}
    \\
    L_{Quality}=\frac{1}{N}\sum\limits_{n=1}^N [(Q_n-\hat{Q}_n)^2+\frac{\alpha_Q}{F_n}\sum\limits_{l=1}^{F_n}(Q_n-\hat{q}_{nl})^2]
    \\
    L_{Intelligibility}=\frac{1}{N}\sum\limits_{n=1}^N [(I_n-\hat{I}_n)^2+\frac{\alpha_I}{F_n}\sum\limits_{l=1}^{F_n}(I_n-\hat{i}_{nl})^2]
    \\
    \end{array} 
\end{equation}
where ${Q_n}$ and ${I_n}$ represent the actual Quality and Intelligibility scores of the $n$-th training utterance, respectively. $\hat{Q}_n$ and $\hat{I}_n$ are the predicted Quality and Intelligibility scores of the $n$-th training utterance. The total number of training utterances is denoted by $N$. ${F_n}$ represents the total number of frames in the $n$-th training utterance, which is the sum of the number of frames of the PS, LFB, and WS features. $\hat{q}_{nl}$ and $\hat{i}_{nl}$ are the predicted frame-level Quality and Intelligibility scores of the $l$-th frame of the $n$-th training utterance, respectively. The weights between utterance-level and frame-level losses are determined by $\alpha_Q$ and $\alpha_I$, while the weights between Quality and Intelligibility are determined by $\gamma_{1}$ and $\gamma_{2}$. 

\graphicspath{ {./images/} }
\begin{figure}[t]
\centering
\includegraphics[width=8cm]{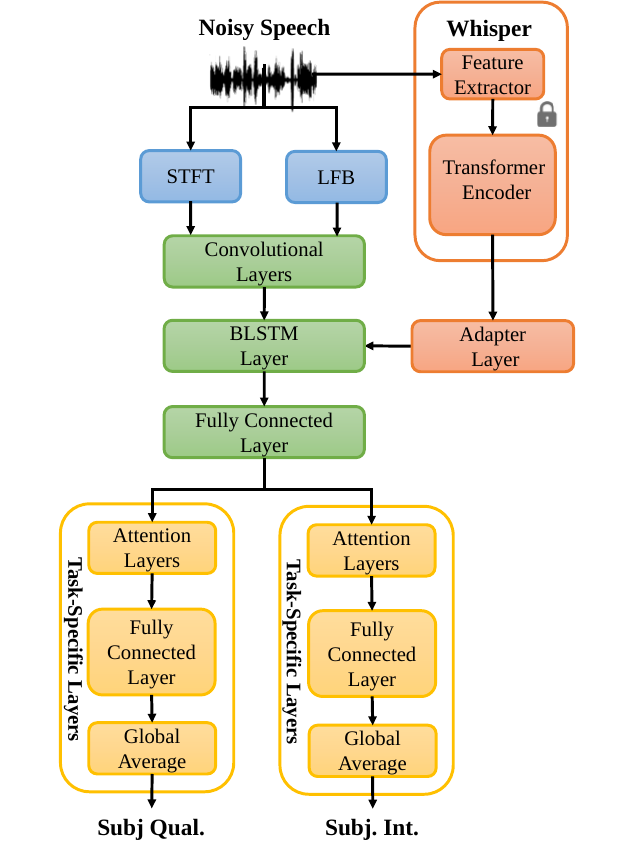} 
\caption{Architecture of the MOSA-Net+ model.} 
\label{fig:mosanet}
\end{figure}

\section{Whisper's and SSL's Embedding Analysis}
To provide additional information about the characteristics of the embedding features of Whisper and SSLs, we aim to estimate the distance score between the noisy/enhanced embedding and its corresponding ground-truth embedding. The embedding features can be generated in the following manner:

\begin{equation}
\label{eq:loss4}
    \begin{array}{c}
\Tilde{\textbf{X}}_{mod}=E_{mod}(F_{mod}(\bold{X}))
\\
\Tilde{\textbf{Y}}_{mod}=E_{mod}(F_{mod}(\bold{Y}))

    \end{array} 
\end{equation}
where $\Tilde{\textbf{X}}_{mod}$ and $\Tilde{\textbf{Y}}_{mod}$ are extracted clean and noisy/enhanced embedding of Whisper or SSL, respectively, with a time dimension $T$ and feature dimension $F$. $F_{mod}$ and $E_{mod}$ are the feature extraction layer and encoder layer of Whisper or SSL, respectively. Afterward, the mean square error (MSE) is used to compute the distance between $\Tilde{\textbf{X}}_{mod}$ and $\Tilde{\textbf{Y}}_{mod}$ in the following manner:

\begin{equation}
\label{eq:loss5}
d_{mod}=\frac{1}{TF} \sum\limits_{t,f}^{T,F}(\Tilde{\textbf{X}}_{mod}[t,f]-\Tilde{\textbf{Y}}_{mod}[t,f])^2
\end{equation}
Then, we calculate the correlation score between the estimated distances, $d_{mod}$, derived from the Whisper and SSL models. We assume that a higher correlation indicates greater similarity in how the Whisper and SSL models capture information from the given audio waveform.

\section{Experiments}
\subsection{Experimental Setup}
The MOSA-Net+ model was tested on the TMHINT-QI dataset \cite{TMINT-QI} and noisy-and-enhanced track of VoiceMOS Challenge 2023 \cite{cooper2023voicemos} \footnote{ \href{https://github.com/dhimasryan/TMHINT-QI_VoiceMOS2023}{The link for the audio files and ground truth scores for the VoiceMOS Challenge 2023 - Track 3 (Noisy-and-Enhanced Track).}}. TMHINT-QI dataset contains clean, noisy, and enhanced speech samples from five different SE systems (Karhunen-Loeve Transform (KLT) \cite{klt}, Minimum-mean Square Error (MMSE) \cite{mmse}, Fully Convolutional Network (FCN) \cite{FCN}, Deep Denoising Autoencoder (DDAE) \cite{DDAE}, and Transformer \cite{Trans}). The dataset was evaluated by 226 listeners who rated both the quality and intelligibility of 108 utterances. The quality score ranges from 1 to 5, while the intelligibility score ranges from 0 to 1. For training, 15,000 utterances were selected, each evaluated by one listener, and for testing, 1,900 utterances were selected, each evaluated by 2 to 3 listeners.

The noisy-and-enhanced track of VoiceMOS Challenge 2023  also utilizes the TMHINT-QI dataset. In this track, a new split between training and validation data has been adopted. Furthermore, an additional listening test was conducted to guarantee that each utterance received evaluations from at least two listeners. The training set encompasses 11,053 utterances, incorporating clean, noisy, and four speech enhancement systems: MMSE, DDAE, FCN, and Transformer.

During the evaluation phase, our main focus is on assessing both unseen noise types and speech enhancement systems. The evaluation set consists of noisy, clean, and enhanced utterances, covering three seen noise conditions (babble, white, and pink noises) and one unseen noise condition (street). Additionally, it includes three seen enhanced systems (FCN, MMSE, Transformer) and introduces two new, unseen enhanced systems: the Conformer-based Metric Generative Adversarial Network (CMGAN) \cite{cao22_interspeech} and DEMUCS \cite{défossez2021music}. In total, the test set comprises 1,960 utterances.

Three evaluation metrics, namely mean square error (MSE), linear correlation coefficient (LCC), and Spearman's rank correlation coefficient (SRCC) \cite{srcc}, were used to measure the performance of MOSA-Net+. Lower MSE values indicate better predictions, while higher LCC and SRCC values indicate a stronger correlation between predicted and ground-truth scores.

\begin{table}[ht]
\caption{Detailed configuration of pre-trained models}
\footnotesize
\begin{center}
 \begin{tabular}{c||c||c||c||c} 
 \hline
 \hline
  \multirow{2}{*}{\textbf{Model}} & \multirow{2}{*}{\textbf{Params.}} & 
  \multirow{2}{*}{\textbf{Dataset size}} &
  \multirow{2}{*}{\textbf{Language}} &
  \multirow{2}{*}{\textbf{FT}}   
  \\
& & &\\
 \hline
 \hline
W2V&95M&60k&English&Yes\\ \hline 
HuBERT&316M&60k&English&No\\ \hline
Whisper&769M&680k&Multilingual&No\\ \hline
MMS&1000M&500k&Multilingual&No\\ \hline

\end{tabular}
\end{center}
\end{table}

\begin{table}[t]
\caption{LCC, SRCC, and MSE from MOSA-Net+ with different latent representation from HuBERT, W2V, MMS, and Whisper for human listening test prediction.}
\footnotesize
\begin{center}
 \begin{tabular}{c||c||c||c} 
 \hline
 \hline
 \textbf{Model} &\textbf{LCC} & \textbf{SRCC} & \textbf{MSE}  \\ [0.5ex] \cline{2-4}
 \hline\hline
  \multicolumn{4}{c} {Speech Quality Prediction}
\\ \hline
HuBERT&0.777&0.724&0.411\\\hline
W2V&0.804&0.758&0.360\\\hline
MMS&0.811&0.766&0.362\\\hline
Whisper&0.815&0.776&0.344\\\hline
Whisper+MMS&\textbf{0.816}&0.777&0.344\\\hline
Whisper+ W2V&\textbf{0.816}&\textbf{0.778}&\textbf{0.343}\\\hline
 \hline
  \multicolumn{4}{c} {Speech Intelligibility Prediction} \\
 \hline
HuBERT&0.740&0.698&0.023\\\hline
W2V&0.796&0.712&0.018\\\hline
MMS&\textbf{0.809}&0.732&0.018\\\hline
Whisper&0.807&\textbf{0.738}&\textbf{0.017}\\\hline
Whisper+MMS&0.785&0.744&0.020\\\hline
Whisper+W2V&0.807&0.733&\textbf{0.017}\\\hline
 \hline

\end{tabular}
\end{center}
\end{table}

\subsection{TMHINT-QI Experimental Results}
\subsubsection{Whisper for Speech Assessment Model}
Our study aims to assess Whisper's suitability in generating latent representation features for deploying the MOSA-Net+ model. We compare it with MMS, which utilizes an additional linear layer of CTC decoder for character mapping, and two SSL models: HuBERT (non-fine-tuned) and Wav2Vec 2.0 (W2V) (fine-tuned). Prior research (\cite{mosa-net, ssl-mos}) found that HuBERT outperformed non-fine-tuned W2V in speech assessment tasks. Conversely, fine-tuned W2V exhibited more robustness compared to its fine-tuned HuBERT counterpart \cite{ssl-mos}. For detailed configuration specifics of the pre-trained models, please refer to Table 1.

The MOSA-Net+ model's training parameters were configured to extract the PS features, and each speech waveform was converted into a 257-dimensional spectrogram using a 512-point STFT with a Hamming window of 32 ms and a hop of 16 ms. The speech waveform was then processed using the LFB and either SSL, MMS or WS model. The output of the PS and LFB feature concatenation was then mapped to 12 convolutional layers, with four channels each (16, 32, 64, and 128). The output of the convolutional layer was then concatenated with the extracted features from the SSL/Whisper model and processed using a one-layered BLSTM (with 128 nodes) and a fully connected layer (with 128 neurons). Two different branches consisting of an attention layer, a fully connected layer (with one neuron), and a global average operation were used to generate the predicted quality and intelligibility scores, respectively. In addition, we set $\gamma_{1}=1$, $\gamma_{2}=1$, and 0.00001 for the learning rate.

As shown in Table 2, we observe that both MMS and Whisper consistently outperformed HuBERT and W2V with relatively higher correlation scores for quality and intelligibility prediction. These results suggest the advantage of using labeled datasets while deploying the speech representation model. Interestingly, compared to W2V, MMS and Whisper do not require an additional fine-tuning process and can maintain a robust speech representation. Moreover, while comparing MMS and Whisper, Whisper can achieve overall better prediction performance than the MMS, which may be due to the advantages of larger training data for deploying the model, despite that the model size of Whisper is smaller than MMS.

Based on the promising performance achieved by Whisper for deploying cross-domain features to train MOSA-Net+ model, we intend to analyze how much improvement can be achieved if we concatenate the acoustic features from "Whisper and W2V" and "Whisper and MMS". We utilize Eq. \ref{eq:loss4} and Eq. \ref{eq:loss5} to derive the estimated distance $d_{mod}$. Subsequently, we employ $d_{mod}$ to compute the correlation matrix, as depicted in Fig. 2. As shown in Table 2, combining "Whisper and SSL model (W2V)" or "Whisper and MMS" leads to a minor boost in prediction performance, with the improvement being rather modest. In addition, the experimental results from Table 2 and Fig. 2 show an inverse relationship between feature correlation and performance. The results indicate that combining features with higher correlation tends to result in lower performance. Specifically, in the context of 'Whisper and W2V' versus 'Whisper and MMS', it appears that 'Whisper and W2V' yielded better performance compared to the 'Whisper and MMS'. Moreover, by considering the computation cost and overall prediction performance, the use of Whisper without additional concatenation from the MMS or other SSL models is already a decent configuration for deploying MOSA-Net+ model, as we can reduce computation time to perform additional fine-tuning or additional feature extraction process while maintaining overall satisfactory prediction performance.

\graphicspath{ {./images/} }
\begin{figure}[t]
\centering
\includegraphics[width=7.5cm]{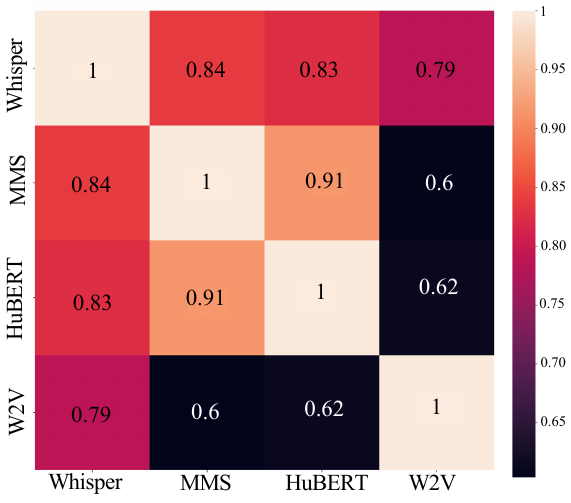} 
\caption{Correlation analysis of the embedding features between Whisper and SSL models.} 
\label{fig:heat}
\end{figure}

\begin{table}[ht]
\caption{LCC, SRCC, and MSE results between Intrusive Methods, MOS-SSL, MOSA-Net, MOSA-Net+, and MOSA-Net+$_{adapt}$ for Human Listening test prediction.}
\footnotesize
\begin{center}
 \begin{tabular}{c||c||c||c} 
 \hline
 \hline
 \textbf{Model} &\textbf{LCC} & \textbf{SRCC} & \textbf{MSE}  \\ [0.5ex] \cline{2-4}
 \hline\hline
  \multicolumn{4}{c} {Speech Quality Prediction}
\\ \hline
CSIG\cite{hu2007evaluation}&0.555&0.453&-\\\hline
CBAK\cite{hu2007evaluation}&0.545&0.343&-\\\hline
COVL\cite{hu2007evaluation}&0.556&0.450&-\\\hline
MOS-SSL \cite{ssl-mos}&0.787&0.746&0.440\\ \hline
MOSA-Net \cite{mosa-net} &0.805&0.763&0.356\\ \hline
MOSA-Net+ &\textbf{0.815}&\textbf{0.776}&\textbf{0.344}\\ \hline
MOSA-Net+$_{adapt}$ &0.812&0.774&0.346\\ \hline
 \hline
  \multicolumn{4}{c} {Speech Intelligibility Prediction} \\
 \hline
ESTOI \cite{estoi} &0.461&0.465&0.162\\ \hline
MOS-SSL \cite{ssl-mos}&0.760&0.655&0.024\\ \hline
MOSA-Net\cite{mosa-net}&0.807&0.730&\textbf{0.017}\\ \hline
MOSA-Net+ &0.807&0.738&\textbf{0.017}\\ \hline
MOSA-Net+$_{adapt}$&\textbf{0.818}&\textbf{0.745}&\textbf{0.017}\\ \hline
 \hline

\end{tabular}
\end{center}
\end{table}

\begin{table*}[t]
\caption{Performance evaluation on noisy-and-enhanced track of VoiceMOS Challenge 2023.}
\footnotesize
\begin{center}
 \begin{tabular}{c||c||c||c||c||c||c||c||c||c} 
 \hline
 \hline
 \textbf{Team ID} &\textbf{Systems} &\textbf{UTT-MSE} & \textbf{UTT-LCC} & \textbf{UTT-SRCC} &\textbf{UTT-KTAU} & \textbf{SYS-MSE} & \textbf{SYS-LCC}& \textbf{SYS-SRCC}& \textbf{SYS-KTAU}  \\ [0.5ex] \cline{2-10}
 \hline\hline
T02&MOSA-Net+&\textbf{0.343}&\textbf{0.803}&\textbf{0.780}&\textbf{0.594}&\textbf{0.082}&\textbf{0.952}&\textbf{0.956}&\textbf{0.828}\\\hline
T06&LE-SSL\cite{qi2023lesslmos}&0.688&0.684&0.636&0.475&0.404&0.769&0.749&0.635\\\hline
T10&-&1.034&0.548&0.520&0.374&0.483&0.755&0.740&0.589\\\hline
T09&KAQ\cite{kaq}&0.586&0.583&0.566&0.403&0.248&0.690&0.720&0.569\\\hline
T04&-&0.885&0.665&0.617&0.456&0.590&0.752&0.715&0.590\\\hline
T08&-&0.999&0.584&0.478&0.338&0.628&0.745&0.621&0.454\\\hline
B02&UTMOS\cite{saeki22c_interspeech}&2.126&0.611&0.477&0.338&1.763&0.769&0.621&0.466\\\hline
B01&SSLMOS\cite{ssl-mos}&3.356&0.518&0.403&0.280&2.986&0.637&0.487&0.357\\\hline
T01&-&3.434&0.511&0.396&0.274&3.056&0.635&0.486&0.355\\\hline
\end{tabular}
\end{center}
\end{table*}

\subsubsection{Comparison with other Methods}
In the next experiment, we aim to compare the performance of MOSA-Net+ with two SSL-based speech assessment models: (1) MOSA-Net \cite{mosa-net}: the original version of MOSA-Net+, which utilizes cross-domain features (PS+LFB+FT-SSL) and weight initialization from MOSA-Net trained on objective scores such as PESQ, STOI, and SDI; (2) MOS-SSL\cite{ssl-mos}: a model that fine-tunes wav2vec 2.0 to predict MOS scores. This is done by mean-pooling the model's output embedding and adding a linear output layer on top of it. We also chose various intrusive speech quality prediction methods, such as CSIG, CBAK, and COVL \cite{hu2007evaluation} as well as the ESTOI \cite{estoi}. Along with that, we also deployed MOSA-Net+$_{adapt}$: same as MOSA-Net+ model, except that the weight is initialized using MOSA-Net model \cite{mosa-net}, which was trained on objective assessment metrics (PESQ, STOI, and SDI).

The results in Table 3 consistently demonstrate MOSA-Net+'s superior performance in all evaluation metrics over the other systems, which confirms the benefits of using Whisper to develop a robust speech representation for the MOSA-Net+ model. Interestingly, MOSA-Net+$_{adapt}$ can notably enhance the accuracy of predicting subjective intelligibility scores, confirming the advantages of the knowledge transfer mechanism. Finally, this experiment provides further evidence that Whisper can achieve decent speech representation to improve the speech assessment model's performance in a zero-shot manner without requiring an online fine-tuning process.

\subsection{VoiceMOS Challenge 2023 Experimental Results}
Following the satisfactory performance achieved by MOSA-Net+ in the previous experiments,  we next evaluate MOSA-Net+ on VoiceMOS Challenge 2023. For VoiceMOS Challenge 2023, MOSA-Net+ is exclusively trained using the noisy-and-enhanced track provided by the organizing committee. The goal of the noisy-and-enhanced track is to estimate the mean opinion score (MOS) of the quality score. Therefore, we selected MOSA-Net+ without domain adaptation to deploy the model, considering its best performance in the previous experiments. In detail, the setup involves employing cross-domain features, specifically a combination of PS+LFB+WS, as the acoustic features. The model architecture selected for training is CNN-BLSTM with an attention mechanism, and a multi-task model architecture is also utilized during the training phase. In this context, the model is trained using both MOS and intelligibility scores as labels, following the objective function defined in Eq. 1. However, during inference, we use the model to estimate the MOS score. In addition, we set $\gamma_{1}=1$, $\gamma_{2}=1$, and 0.00001 for the learning rate.

In Table 4 \footnote{The evaluation for ranking was performed by the VoiceMOS 2023 Committee. \cite{cooper2023voicemos}}, MOSA-Net+ exhibits superior performance compared to LE-SSL-MOS employing SSL fine-tuning with listener embedding, KAQ utilizing a stacking process, four other teams, and two baseline systems (UTMOS and SSL-MOS), showcasing a notable margin of improvement in all evaluation metrics. In addition, unlike the other mentioned systems (LE-SSL-MOS, UTMOS, and SSL-MOS), MOSA-Net+ is the only system that employs the Whisper model to generate the acoustic feature, whereas the other systems use SSL to generate the acoustic features. Therefore, it reaffirms the advantages of Whisper to provide decent acoustic features for better prediction capability of a non-intrusive speech assessment model.



\graphicspath{ {./images/} }
\begin{figure}[t]
\centering
\includegraphics[width=8.5cm]{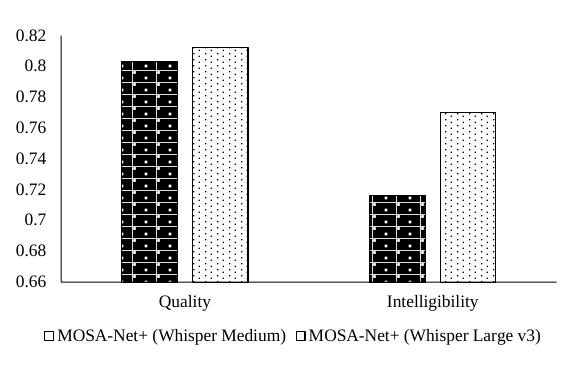} 
\caption{Performance comparison between MOSA-Net+ (Whisper Medium) and MOSA-Net+(Whisper Large v3).} 
\label{fig:heat}
\end{figure}

\subsection{Comparsion of Different Versions of Whisper}
In this section, we intend to further confirm the advantages of Whisper acoustic features by comparing the latest version of the Whisper model with the previous version model. In our previous experiments, we used the Whisper medium to deploy the systems, where our systems already achieved state-of-the-art performance in the noisy-and-enhanced track of the VoiceMOS Challenge 2023. With the release of Whisper large v3, which employed a larger mel frequency bin of the input features and employed an additional language training set, we again test and compare the performance of the model. As shown in Fig. 3, consistent performance improvement is achieved by the MOSA-Net+ model employing Whisper large v3 features. This further indicates the advantages of the latest Whisper model in providing more representative acoustic features for robust speech assessment performance.

\section{Conclusions}
This paper presents MOSA-Net+, an improved version of MOSA-NET that predicts human-based speech quality and intelligibility. MOSA-Net+ uses a well-known weakly supervised model (Whisper) to generate cross-domain features. The model employs a CNN-BLSTM architecture with an attention mechanism and is trained using a multi-task learning approach to predict subjective listening test scores. Experimental results show that incorporating Whisper's embedding features notably improves the robustness of MOSA-Net+. Additionally, combining the embedding features from Whisper and SSL models only results in a marginal improvement. Furthermore, when evaluated on the TMHINT-QI dataset, MOSA-Net+ outperforms MOSA-Net, MOS-SSL, and several intrusive metrics in all evaluation metrics for predicting quality and intelligibility scores. Finally, in the noisy-and-enhanced track of VoiceMOS Challenge 2023, MOSA-Net+ can achieve the best performance among nine systems. In the future, we plan to explore the potential of Whisper in developing a robust speech assessment model for more unseen tasks. Meanwhile, we will also explore a direct integration of the speech assessment model with speech processing applications.

\bibliographystyle{IEEEbib}
\bibliography{refs}
\end{document}